\documentclass[aps,prl,10pt,twocolumn,amsmath,amssymb,floatfix,superscriptaddress,longbibliography]{revtex4-2}
\usepackage{amsmath}
\usepackage{graphicx} 
\usepackage{bm} 
\usepackage{hyperref}
\hypersetup{
    colorlinks,
    citecolor=black,
    filecolor=black,
    linkcolor=black,
    urlcolor=black
}
\usepackage{braket,bbold,color,enumerate}
\usepackage{cleveref}

\newcommand{\bs}[1]{\boldsymbol{#1}}

\newcommand{\SSH}{\mathrm{SSH}}
\usepackage{xcolor}

\begin{document}

\title{Topological Dislocation Response in Elementary Semiconductors}

\author{Yuteng Zhou}
\affiliation{Blackett Laboratory, Imperial College London, London SW7 2AZ, United Kingdom}

\author{Alexandre Chaduteau}
\affiliation{Blackett Laboratory, Imperial College London, London SW7 2AZ, United Kingdom}

\author{Frank Schindler}
\affiliation{Blackett Laboratory, Imperial College London, London SW7 2AZ, United Kingdom}

\begin{abstract}
We study elementary semiconductors and insulators that are symmetric under spatial inversion: silicon, diamond, germanium, and black phosphorene. These materials are ideal candidates for realizing obstructed atomic insulators, which differ from trivial atomic insulators by a quantized spatial shift of their electronic Wannier centers with respect to the atomic lattice. We use symmetry indicator invariants that allow the prediction of non-trivial responses to crystal dislocations with integer Burgers vector in these materials, and generally, in all inversion-symmetric obstructed atomic insulators. Unlike weak topological insulators where the response only depends on the Burgers vector, the dislocation response of 3D inversion-symmetric obstructed atomic insulators can also be affected by the line vector of the dislocation. In such insulators, we find that edge dislocations generically exhibit a non-trivial response, while in all 3D obstructed atomic insulators, screw dislocations always display a trivial response. With the aid of numerical simulations of realistic tight-binding models, we confirm the presence of mid-gap polarization bands localized along dislocations in silicon, diamond, and germanium.
\end{abstract}

\maketitle
\emph{Introduction---}
The theory of non-interacting topological insulators (TIs)~\cite{RevModPhys.89.041004, RevModPhys.82.3045, Vergniory_2019} can be considered quite mature. TIs can host edge states that are robust under perturbations; their presence is diagnosed by bulk topological invariants. Besides edges, crystals will inevitably host imperfections such as crystal defects~\cite{Boer2023} that alter the local geometry of the lattice. Refs.~\onlinecite{Ran2009, teo_kane_defects_in_TIs_TSCs, PhysRevLett.108.106403, Queiroz19, Schindler_2022, Disloc1,Disloc2,Disloc3,Disloc4,Disloc5,Disloc6,Disloc7,Disloc8,Disloc9,Disloc10,Disloc11,Disloc12} have defined topological invariants to diagnose the presence of robust bound states localized at such defects. In particular it was found that dislocations -- stable line defects in three-dimensional (3D) crystals that disrupt translational symmetry only locally -- can host topologically protected one-dimensional (1D) ``first-order'' gapless helical modes~\cite{Ran2009} and zero-dimensional (0D) ``higher-order'' end states~\cite{Schindler_2022}, which are protected by spinful time reversal ($\mathcal{T}$) and inversion ($\mathcal{I}$) symmetry, respectively. This response has been discussed for integer Burgers vector dislocations in insulators such as BiSb~\cite{Ran2009}, PbTe monolayers and SnTe~\cite{Schindler_2022}, elementary bismuth~\cite{Nayak2019}, as well as in acoustic and photonic systems~\cite{Lin2023Topological, Ye2022}.

We here apply this theory to some of the most widely studied $\mathcal{I}$- and $\mathcal{T}$-symmetric elementary semiconductors~\footnote{Others include: $\alpha$-Sn, with near-zero band gap and potential band inversions induced by strain at the $\Gamma$ point, which does not contribute to a dislocation response; and boron, whose tight-binding model has a complex structure and large unit cell, rendering the study of dislocation-bound states significantly more involved.}: 3D silicon, diamond, germanium, and two-dimensional (2D) black phosphorene (BP). These materials are obstructed atomic insulators (OAIs)~\cite{Bradlyn_2017, liu2024massive1ddiracline, PhysRevB.105.165135, Elcoro_2021,OAI1,OAI2,OAI3,OAI4,OAI5,OAI6,OAI7,OAI8,OAI9,OAI10,OAI11,OAI12,OAI13,OAI14,OAI15,OAI16,OAI17,OAI18,OAI19,OAI20} whose symmetric and exponentially-localized electronic Wannier functions are not centered at atomic sites~\cite{Wannier_original}. We use topological invariants inspired by Ref.~\onlinecite{Schindler_2022} to diagnose the dislocation response of 2D BP and 3D silicon, diamond and germanium. Via realistic tight-binding models, we confirm that when introducing edge dislocations, these semiconductors exhibit floating bands localized along the dislocation inside the gap of the energy spectrum~\cite{Yuteng_Data_2026}. We call them polarization bands as they arise from the bulk polarization, similar to the end states of the Su-Schrieffer-Heeger (SSH) chain~\cite{PhysRevLett.42.1698,PhysRevB.22.2099}. Since it is impossible to remove the degeneracy between polarization bands associated with spatially separated (and $\mathcal{I}$-related) dislocations without breaking $\mathcal{I}$ and $\mathcal{T}$ symmetry, a filling anomaly occurs and furnishes a topologically protected defect response of the OAI phase~\cite{Benalcazar2019,Filling1}. We also prove that screw dislocations cannot bind polarization bands in all 3D OAIs.

\emph{Topological invariants for dislocation response---} In 2D TIs with $\mathcal{I}$ and $\mathcal{T}$ symmetry, the $\mathbb{Z}_2$ Fu-Kane invariant~\cite{fu_kane_mele_TIs_in_3D, fu_kane_TIs_with_inversion} diagnoses the presence of an odd number of robust pairs of helical edge states. This notion was extended to 3D $\mathcal{I}$- and $\mathcal{T}$-symmetric TIs which now have four $\mathbb{Z}_2$ invariants $(\nu_0, \nu_1, \nu_2, \nu_3)$ implying the presence of topological surface states. Here $\nu_i \in \{0,1\}$, $i = 1,2,3$, are called weak invariants, which may be packaged into a vector $\bs{M}_\nu=(\nu_1\bs{b}_1+ \nu_2\bs{b}_2+\nu_3\bs{b}_3)/2$, where $(\bs{b}_1,\bs{b}_2,\bs{b}_3)$ are reciprocal basis lattice vectors~\cite{Schindler_2022, Ran2009}. These invariants rely on translational symmetry. The other ``strong'' topological invariant $\nu_0=0,1$ in principle does not require translational invariance (though it may be easier to calculate in momentum space) and describes surface states that are robust against disorder. A 3D weak TI -- which only has the weak invariants non-vanishing -- can be adiabatically related to a disconnected stack of 2D TIs along the real-space direction dual to $\bs{M}_\nu$.

\begin{figure*}[t]
    \centering
    \includegraphics[width=1\linewidth]{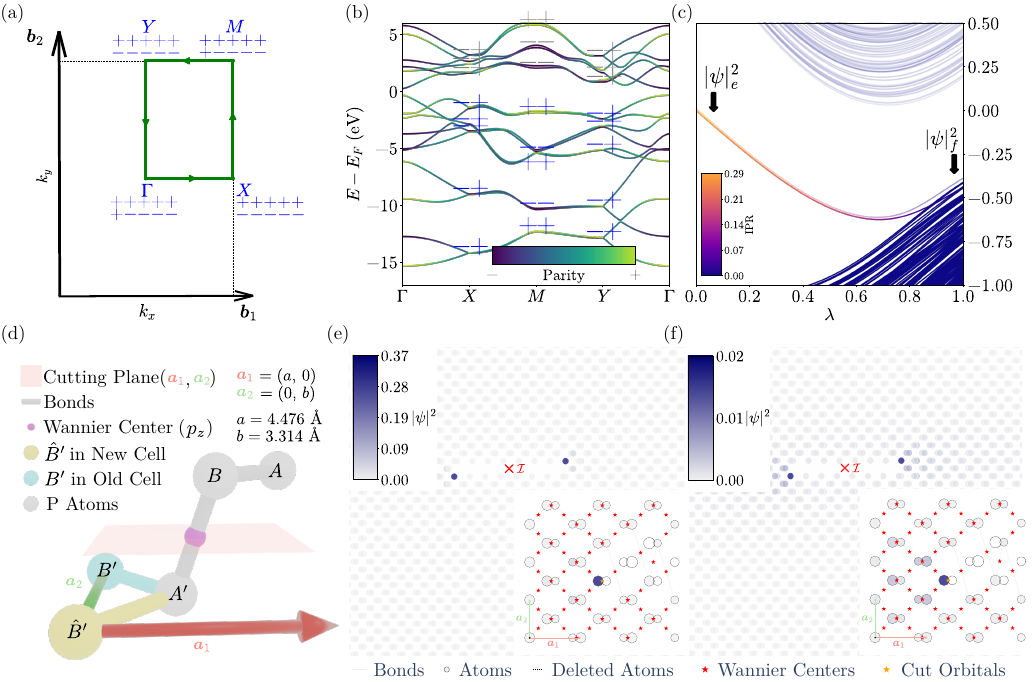}
    \caption{Dislocation response of black phosphorene (BP). (a) Brillouin zone of BP, showing time-reversal invariant momentum (TRIM)~\cite{Schindler_2022} points and their occupied inversion eigenvalues. (b) Band structure along the path shown in (a), with associated inversion eigenvalues at TRIMs. (c) Spectrum in presence of a pair of $\mathcal{I}$-related dislocations, plotted against the interpolation parameter $\lambda$ [see Eq.~\eqref{eq:interpolating_hamiltonian}]. Two polarization bands appear inside the gap. One of them is slightly shifted to aid visualization. We color-code their IPR, which is maximal at $\lambda = 0$ (90° model) and minimal at $\lambda = 1$ (realistic model) [see Eq.~\eqref{eq:interpolating_hamiltonian}]. Simulations are performed on a lattice of $30 \times 30$ unit cells spanned by $\bs{a}_1$ and $\bs{a}_2$. The dislocation is introduced by cutting the lattice (keeping unit cells intact), removing $7$ unit cells, and gluing back together. (d) Unit cell of BP, consisting of $4$ phosphorus atoms, each of which has $5$ valence electrons. The commonly used unit cell consists of the gray and blue atoms ($A$, $B$, $A'$ and $B'$). By contrast, the unit cell we use consists of the gray and yellow atoms ($A$, $B$, $A'$ and $\hat{B}'$). This choice ensures that the edge dislocation preserves $\mathcal{I}$ symmetry. The positions of these atoms in terms of fractional lattice vectors are: $\tau_A = \left[ \left(1/2- u \right)a,\ b/2,\ c \right],\ \tau_B = (ua,\ 0,\ c),\ \tau_{A'} = -\tau_B,\ \tau_{B'} = -\tau_A + \bs{a}_2$ and $\tau_{\hat{B}'} = -\tau_A$, where $u = 0.08056$ and $c = 1.0654\,\text{\AA}$~\cite{BP_structure}. The pink sphere represents the Wannier center induced by the $p_z$ orbital; the red plane, which in our simulation connects the two $\mathcal{I}$-related edge dislocations, cuts through it. (e),(f)~ Real-space wavefunction density $|\psi|^2$ plots of polarization states for $\lambda = 0$ and $\lambda = 1$, respectively, with inversion center shown in red.}
    \label{fig:phosphorene_results}
\end{figure*}

\begin{figure*}[t]
    \centering
    \includegraphics[width=1\linewidth]{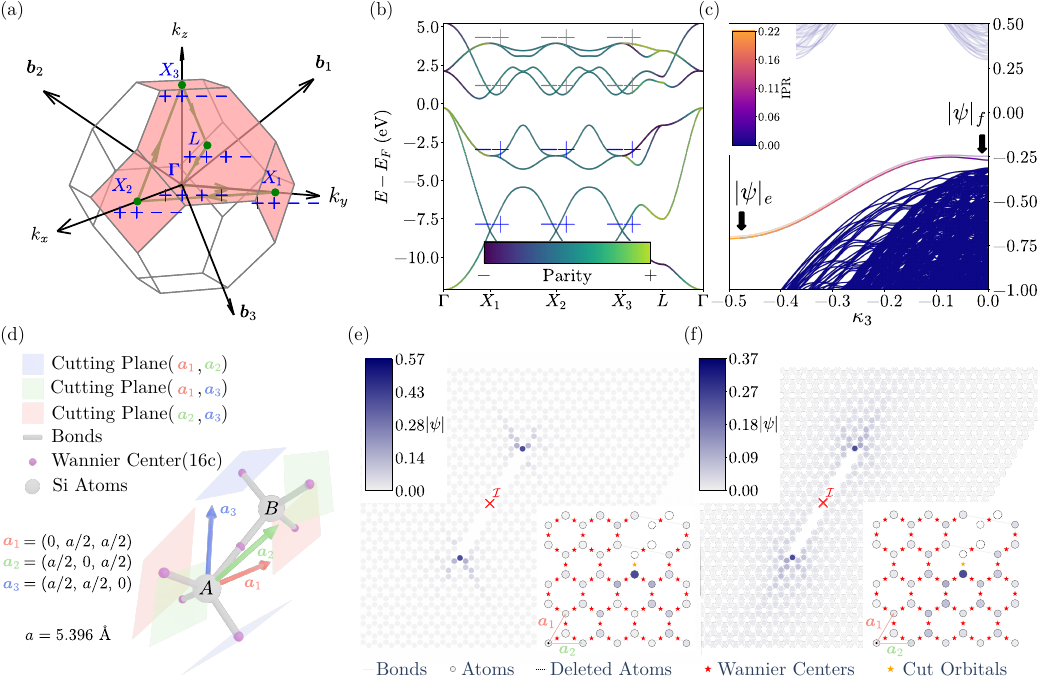}
    \caption{Edge dislocation response of 3D silicon. (a) Brillouin zone of silicon, showing time-reversal invariant momentum (TRIM)~\cite{Schindler_2022} points and their occupied inversion eigenvalues. (b) Band structure along the path shown in (a), with associated inversion eigenvalues at TRIMs. (c) Band structure with a pair of $\mathcal{I}$-related dislocations present, plotted against the only remaining momentum quantum number $\kappa_3=(\bs{k}\cdot{\bs{a}_3})/{2\pi}$. The band structure for positive $\kappa_3$ can be obtained by reflection along the $\kappa_3=0$ axis because of $\mathcal{T}$ symmetry. Simulations are performed on a lattice of $30 \times 30$ unit cells spanned by $\bs{a}_1$ and $\bs{a}_2$ with $100$ sample points for momentum $\kappa_3$. The dislocation is introduced by cutting the 3D material along a plane (keeping unit cells intact), removing $10$ unit cells, and then gluing the lattice back together~\cite{Schindler_2022}. We enforce periodic boundary conditions in all three directions.
    (d) Real-space unit cell geometry, common to all three 3D semiconductors that we study. There are two silicon atoms in a unit cell, with positions $\tau_A =  (0,0,0)$ and $\tau_B = \frac{1}{4}(1, 1, 1)$ in fractional lattice vectors. (e),(f): Real-space wavefunction $|\psi|$ plots for polarization states at $\kappa_3=-1/2$ and $\kappa_3=0$, respectively, with inversion center shown in red. We plot the absolute value of the amplitude instead of the density here for a more detailed visualization.}\label{fig:silicon_results}
\end{figure*}

For OAIs, all four Fu-Kane invariants are trivial by the assumption that we can find exponentially localized Wannier states, so that we do not expect gapless surfaces.
Nevertheless, when introducing dislocations that cut through the Wannier centers, OAIs can host polarization-induced bound states~\cite{Schindler_2022}. More formally, a bulk-defect correspondence can be formalized via the notion of filling anomaly~\cite{Benalcazar2019}: at neutral filling, dislocation bound states that are related by $\mathcal{I}$ symmetry are only half-filled on average. The resulting ground state degeneracy can only be lifted by breaking $\mathcal{I}$ symmetry or closing the bulk gap. In particular, it remains even when the dislocation bound states are pushed out of the gap and hybridize with the bulk states, thus serving as a robust dislocation response diagnosing the OAI phase absent fine tuning. We note that in this case there are no more dislocation-localized eigenstates present in the spectrum, however, an exponentially-localized fractional charge accumulation remains bound to the dislocation~\cite{Benalcazar2019} even when mid-gap states are absent. In ideal situations, furthermore, the defect-localized states form a floating band \emph{inside} the bulk gap. Though this is not guaranteed to happen, we will show that this case is realized in 3D semiconductors.

We now construct topological invariants that predict dislocation-induced filling anomalies in OAIs~\cite{Schindler_2022}.
Consider a generic $d$-dimensional $\mathcal{I}$-symmetric (and optionally $\mathcal{T}$-symmetric) crystal. We label basis lattice vectors by $\bs{a}_i, \;i \in \{1,\ldots d\}$, and reciprocal basis lattice vectors by $\bs{b}_i$. There are $2^d$ time-reversal invariant momentum (TRIM) points in the $d$-dimensional Brillouin zone (BZ). Each point is labeled by $\alpha = (n_1, \dots, n_d)$, a tuple of $d$ numbers $n_i=0,1$ so that its momentum reads $\bs{k}_{\alpha}=\sum_{i=1}^{d}n_i\bs{b}_i/2$. These momenta $\bs{k}_{\alpha}$ satisfy $\bs{k}_{\alpha}=-\bs{k}_{\alpha} \text{ mod } \bs{G}$, where $\bs{G}$ is any linear integer combination of the $\bs{b}_i$. We define the index
\begin{equation}
\delta_\alpha = \prod_{m=1}^{N} \xi_{m}(\bs{k}_{\alpha})\xi^{\text{A}}_{m}(\bs{k}_{\alpha}),
\end{equation}
where $\xi_{m}(\bs{k}_{\alpha})$ and $\xi^{\text{A}}_{m}(\bs{k}_{\alpha})$ are, respectively, the inversion eigenvalues of the occupied bands $m = 1 \dots N$ and for a reference unobstructed atomic limit~\cite{Schindler_2022}, at the TRIM $\bs{k}_{\alpha}$. The reference unobstructed atomic limit has Wannier centers that lie on the atomic sites, with the same lattice structure and unit-cell convention as the corresponding OAI. Topological invariants of a material are defined relative to the reference unobstructed atomic limit, ensuring that the resulting classification is independent of the particular choice of unit cell. The reference unobstructed atomic limits for the materials studied in this work are detailed in the Supplemental Material (SM)~\cite{Note2}.

In 2D, following notation from Ref.~\onlinecite{Schindler_2022}, we define a two-component weak SSH invariant $\bs{M}^{\mathrm{SSH}}_\nu = (\nu^{\mathrm{SSH}}_1\bs{b}_1+\nu^{\mathrm{SSH}}_2\bs{b}_2)/2$
via
\begin{equation}\label{eq: 2D_weak_SSH_invariant}
     \nu_{i}^{\mathrm{SSH}}=\frac12\left[1-\prod_\alpha (\delta_\alpha)^{\bs{k}_{\alpha}\cdot\bs{a}_i/\pi}\right], \quad i\in \{1,2\},
\end{equation}
where $\alpha$ runs over all TRIM points in the first BZ. Intuitively, $\nu_{i}^{\mathrm{SSH}}$ is nonzero and equal to $1$ when the 1D Bloch Hamiltonian defined on the momentum space line $\bs{k}\cdot\bs{a}_i = \pi \mod 2\pi$ has a non-vanishing (time-reversal) polarization~\cite{PhysRevB.47.1651, PhysRevB.48.4442}, as exemplified in an SSH chain~\cite{Alexandradinata14}.
If we consider a pair of dislocations related by $\mathcal{I}$-symmetry,
filling-anomalous bound states will be present when the dislocation Burgers vector $\bs{B}$ satisfies~\cite{Ran2009, Schindler_2022}
\begin{equation}\label{eq:2D_response_condition}
\bs{B}\cdot\bs{M}^{\mathrm{SSH}}_\nu \text{ mod }2\pi = \pi.
\end{equation}
We will use this condition for BP, a 2D semiconductor.

Let us move on to 3D.
Edge dislocations break translational symmetry in two directions out of three, leaving only one good quantum number, a reduced momentum $\kappa_{\parallel}
={\bs{k}\cdot\bs{T}}/{2\pi}$ along the dislocation \emph{line vector} $\bs{T}$. For edge dislocations, $\bs{T}$ is a different lattice vector than the Burgers vector $\bs{B}$. We can therefore view $\kappa_{\parallel}$ as an external pumping parameter between two $\mathcal{I}$-symmetric 2D Hamiltonians, each in the presence of a dislocation, that are defined at $\kappa_{\parallel} = 0$ and $\kappa_{\parallel} = 1/2$. OAIs must have identical weak SSH invariants $\bs{M}_{\nu}^{\mathrm{SSH}}$ for these two 2D Hamiltonians because their 3D weak index $\bs{M}_\nu$ vanishes. As a result, they cannot host the gapless dislocation states of Refs.~\onlinecite{Ran2009, PhysRevB.84.035443}, but they can still stabilize filling-anomalous polarization bands at edge dislocations if $\bs{M}_{\nu}^{\mathrm{SSH}}$ defined at $\kappa_{\parallel} = 0, 1/2$ satisfies Eq.~\eqref{eq:2D_response_condition}.
Conversely, screw dislocations have their line vector $\bs{T}$ parallel to the Burgers vector $\bs{B}$.
At $\kappa_{\parallel} = 0$, the 2D Hamiltonians with or without the screw dislocation are identical~\cite{Schindler_2022} and therefore dislocation-bound polarization bands cannot exist. Since all atomic insulators, obstructed or unobstructed, cannot support topologically protected edge states that connect conduction and valence bands, the response to screw dislocations is always trivial.

We now formalize these observations for general dislocations. Define the 3D weak SSH invariants as a matrix
\begin{equation}\label{eq:M_matrix}
    \hat{M}_\nu^{\mathrm{SSH}} = \frac{1}{4\pi}\sum_{i,j=1}^3\nu^{\SSH}_{ij}\bs{b}_i\otimes\bs{b}_j,
\end{equation}
where $\otimes$ denotes the outer product and
\begin{equation}\label{eq:nu_ij}
     \nu_{ij}^{\mathrm{SSH}}=\frac12\left[1-\prod_\alpha\delta_\alpha^{(\bs{k}_{\alpha}\cdot\bs{a}_i)(\bs{k}_\alpha\cdot\bs{a}_j)/\pi^2}\right].
\end{equation}
Here, $\alpha$ runs over all 8 TRIM points in the 3D BZ.
For any $\mathcal{I}$-symmetric pair of dislocations (edge, screw, or mixed) with Burgers vector $\bs{B}$ and line vector $\bs{T}$, filling-anomalous polarization bands that disperse with a reduced momentum quantum number $\kappa_{\parallel}$ along $\bs{T}$ will be present in a 3D OAI when
\begin{equation}
\label{eq: 3D_response_condition}
\bs{B}\cdot(\hat{M}_\nu^{\mathrm{SSH}}\cdot\bs{T}) = \pi\text{ mod } 2\pi. 
\end{equation}
Eq.~\eqref{eq:nu_ij} implies that $\nu^{\mathrm{SSH}}_{i,j} = \nu^{\mathrm{SSH}}_{j,i}$, meaning the dislocation response is invariant under an exchange of $\bs{B}$ and $\bs{T}$. The trivial response for screw dislocations is reflected in the fact that $\nu^{\mathrm{SSH}}_{i,i}$ reduces to the weak invariant $\nu_{i}$, which is always zero for OAIs.

\emph{Dislocation response of black phosphorene---}
We first consider BP, a 2D material whose puckered honeycomb geometry is depicted in Fig.~\ref{fig:phosphorene_results}(d), which also defines the lattice basis vectors $\bs{a}_1, \bs{a}_2$ and associated lattice constants $a,b$.
The BP Hamiltonian we use interpolates between two models~\cite{PhysRevB.104.125302} with a parameter $0 \leq \lambda \leq 1$. The first model $H_{\text{90°}}(\bs{k})$ assumes a 90° angle between bonds in the puckered structure; it has a chiral symmetry which pins dislocation bound states at zero energy. The second model is a realistic model $H_{\text{black}}(\bs{k})$ of BP, which does not have this additional symmetry. Following Ref.~\onlinecite{PhysRevB.104.125302}, we have
\begin{equation}\label{eq:interpolating_hamiltonian}
    H_\lambda(\bs{k}) = (1-\lambda)H_{\text{90°}}(\bs{k}) + \lambda H_{\text{black}}(\bs{k}).
\end{equation}
Using either model, we obtain the same inversion eigenvalues shown in Figs.~\ref{fig:phosphorene_results}(a) and~(b), which result in $({\nu^{\mathrm{SSH}}_{1}},\,{\nu^{\mathrm{SSH}}_{2}})= (1, \,1)$~\footnote{See the Supplemental Material (containing Refs.~\cite{Nakhaee_2020,PhysRevB.47.1651,fu_kane_TIs_with_inversion,Keating1966,Harrison1980}) for calculations of invariants, as well as a plot summarizing results for diamond and germanium, and a study of the stability of our results to inversion-breaking relaxation.}, and the bulk gap does not close as we tune $\lambda = 0 \rightarrow 1$. Therefore we obtain $\bs{M}^{\mathrm{SSH}}_\nu = \pi(1/a, 1/b)$, indicating a non-trivial response for \emph{any} $\mathcal{I}$-symmetric pair of dislocations with a Burgers vector $\bs{B}$ that is an odd combination of lattice basis vectors. We introduce pairs of $\mathcal{I}$-symmetric dislocations with $\bs{B} = \bs{a}_2=(0,b)$ for BP, satisfying Eq.~\eqref{eq:2D_response_condition}. As summarized in Fig.~\ref{fig:phosphorene_results}, we observe a non-trivial dislocation response, manifested at $\lambda = 0$ in the presence of two exactly overlapping bound states which lie in the middle of the energy gap. Their degeneracy arises from the fact that they are related by $\mathcal{I}$ symmetry and that the defects are located far away from each other.
We plot the full spectrum against $\lambda$ [Fig.~\ref{fig:phosphorene_results}(c)] to see where the bound states end up in the realistic model. As $\lambda \to 1$, although polarization states merge with the valence bands, they still create a state-counting imbalance between conduction and valence bands. This implies a filling anomaly~\cite{Benalcazar2019}, robust under perturbations, that ensures the material is necessarily metallic at $5/8$ filling. 

As a measure of localization, we compute the inverse participation ratio (IPR) for each eigenstate, defined as
\begin{equation}
\mathrm{IPR} = \sum_i \left( \sum_n |\psi_{i,n}|^2 \right)^2  
\end{equation}
where $\psi_{i,n}$ denotes the wavefunction component of orbital $n$ at lattice site $i$. We show the result in Fig.~\ref{fig:phosphorene_results}(c).
The localization length of the dislocation bound states scales inversely with the energy gap as expected~\cite{RevModPhys.66.899, PhysRevB.47.1651}. We plot the corresponding real-space density profile in Figs.~\ref{fig:phosphorene_results}(e) and (f).

\emph{Dislocation response of 3D semiconductors---}
We next discuss silicon, diamond, and germanium, with spatial geometry depicted in Fig.~\ref{fig:silicon_results}(d) and lattice constants $a = 5.396$, $3.596$ and $5.658$ \text{\AA}, respectively. For all three materials, we consider well-known tight-binding Hamiltonians with parameters from Ref.~\onlinecite{Nakhaee_2020} (see also Ref.~\onlinecite{PhysRev.94.1498}). 
In the BZ, the Bloch Hamiltonians restricted to the 1D lines connecting points $X_i$ to $L$ for $i=1, 2, 3$, with inversion eigenvalues shown in Figs.~\ref{fig:silicon_results}(a) and (b), all feature an odd number of band inversions with respect to the reference atomic limit~\footnote{For all materials, we choose $\mathcal{I}$-symmetric unit cells, implying vanishing polarization in the reference atomic limit.}. For all three semiconductors, Eq.~\eqref{eq:nu_ij} then gives $(\nu^{\mathrm{SSH}}_{1,2},\,\nu^{\mathrm{SSH}}_{2,3},\,{\nu^{\mathrm{SSH}}_{3,1}})=(1,\,1,\,1)$, as calculated in the SM~\cite{Note2}. These invariants imply that Eq.~\eqref{eq: 3D_response_condition} is satisfied for \emph{any} pair of \emph{edge} dislocations when $\bs{B}$ and $\bs{T}$ are lattice basis vectors. We introduce pairs of $\mathcal{I}$-symmetric dislocations with $\bs{B} = \bs{a}_1=a/2(0,1,1)$ and $\bs{T}=\bs{a}_3=a/2(1,1,0)$, satisfying Eq.~\eqref{eq: 3D_response_condition} and find that mid-gap dislocation bands are present even at realistic model parameters.
We summarize our results in Fig.~\ref{fig:silicon_results} for silicon, and relegate similar plots for diamond and germanium to the SM.
As in Fig.~\ref{fig:phosphorene_results}, we use the IPR to study the localization properties of the dislocation bound states. The IPR is smallest at $\kappa_3 ={\bs{k}\cdot\bs{a}_3}/{2\pi}= 0$ and largest at $\kappa_3 = 1/2$. We show the real-space density profile of the dislocation states in Figs.~\ref{fig:silicon_results}(e) and (f). At half-filling, there is a filling anomaly \emph{per momentum} $\kappa_3$~\cite{Schindler_2022}. 
As detailed in the SM~\cite{Note2}, we study inversion-breaking relaxations, as well as different boundary terminations. The former can lift the degeneracy of the bound states. The latter creates an imbalance of density between the two defect cores. At this point, there is in general no topological guarantee of polarization bands or a filling anomaly. However, if the deviations from the symmetric limit are small, we still expect remnants of the topological response to survive in realistic samples. Indeed, we show in the SM that while the polarization and visibility of the dislocation states are reduced under realistic conditions, they generically remain in the gap and remain localized throughout parts of momentum space.

\emph{Discussion---}
Using well-known tight-binding models, we have demonstrated nontrivial dislocation responses in a 2D OAI material and three 3D materials that realize the same OAI phase and therefore the same response. One main physical distinction between these materials is their various band gaps and polarization band dispersions, causing varying degrees of localization of the polarization band states, as seen in their IPR. In particular, we have found that the large band gaps of diamond and silicon host cleanly isolated mid-gap dislocation bands that can be probed in experiment~\cite{Nayak2019}, while the bulk bands of germanium and BP merge with the polarization states, so that in general only a filling anomaly remains as a clear-cut topological response in these materials. More generally, we have shown that 3D OAIs host richer potential dislocation responses than their 2D counterparts due to the freedom in choosing a line vector for dislocations in 3D. In 3D OAIs, the response can be non-trivial for edge dislocations but is always trivial for screw dislocations. Notably, this dependence on a line vector is in stark contrast to the case of 3D weak topological insulators, for which the dislocation response depends only on the Burgers vector $\bs{B}$~\cite{Ran2009}. A clear next step would be to extend our theory beyond $\mathcal{I}$-symmetry to materials with other crystalline symmetries.

\emph{Acknowledgments---}
We thank Malcolm Connolly for early discussions on this idea. F.S. thanks Benjamin Wieder, Andrei Bernevig and Titus Neupert for a previous collaboration on a similar topic. A.C. acknowledges support from Imperial College London via a President’s PhD Scholarship. This work was supported by a UKRI Future Leaders Fellowship MR/Y017331/1.

\bibliography{refs}

\end{document}


\title{Supplemental Material for \\``Topological Dislocation Response in Elementary Semiconductors''}

\author{Yuteng Zhou}
\affiliation{Blackett Laboratory, Imperial College London, London SW7 2AZ, United Kingdom}

\author{Alexandre Chaduteau}
\affiliation{Blackett Laboratory, Imperial College London, London SW7 2AZ, United Kingdom}

\author{Frank Schindler}
\affiliation{Blackett Laboratory, Imperial College London, London SW7 2AZ, United Kingdom}

\maketitle
\section{Reference unobstructed atomic limits}
A reference unobstructed atomic limit is chosen by turning off the hopping parameters between orbitals~\cite{Nakhaee_2020}. As a result, the Wannier centers are located at atomic positions and completely isolated from each other, which is the simplest form of an insulator. For example, for silicon, diamond and germanium, there are four orbitals ($s$, $p_x$, $p_y$ and $p_z$) and two atomic positions in each unit cell, which naturally gives $8$ bands. Due to inversion symmetry, the on-site energies for the same orbital at two atomic positions must be the same. We also have four doubly degenerate bands with parity eigenvalues $\xi^{\text{A}}_{m}(\bs{k}_{\alpha}) =\pm 1$ at each time-reversal-invariant momentum (TRIM) in Eq. (1) in the main text (using the choice of an inversion-symmetric unit cell). Regardless of the values of these four on-site energies, the product of all the parity eigenvalues of the occupied bands at each TRIM is $+1$. 

Here, the choice of inversion-symmetric unit cell has an advantage with regard to the Berry phase $\gamma$. According to the modern theory of polarization~\cite{PhysRevB.47.1651}, $\gamma$ is proportional to the polarization (or electric dipole moment) $P$ according to 
\begin{equation}
    P = -\frac{e}{2\pi}\gamma \quad \mathrm{mod}\ e,
\end{equation}
where $e$ is the elementary electron charge. In an inversion-symmetric system, the total Berry phase $\gamma$ for all $N$ occupied bands along the direction containing two TRIMs $\bs{k}_{\alpha}$ and $\bs{k}_{\beta}$ can be calculated through~\cite{fu_kane_TIs_with_inversion}
\begin{equation}
    \gamma
    =
    \frac{\pi}{2}
    \left[
    1
    -
    \prod_{m=1}^{N}
    \xi_m(\bs{k}_{\alpha})
    \xi_m(\bs{k}_{\beta})
    \right].
\end{equation}
When electrons lie at the atomic positions, $P=0$, so $\gamma$ must be zero along any direction. This means any path connecting two TRIMs has a total product of occupied parity eigenvalues of $+1$ for these two points. This simplifies the calculation in Eq.~(1) of the main text by converting the product $\prod_{m=1}^{N}\xi^{\text{A}}_{m}(\bs{k}_{\alpha})$ into a single factor of $+1$ or $-1$ for all TRIMs. Also then $\delta_\alpha$ becomes
\begin{equation}
\delta_\alpha = \pm\prod_{m=1}^{N} \xi_{m} (\bs{k}_\alpha).
\end{equation}
Since all invariants depend on the product of $\delta_\alpha$ of all the TRIMs, the $\pm$ sign introduced by the reference OAI using an inversion-symmetric unit cell does not change them.
If unobstructed atomic limits are instead chosen to hold a non-zero $P$ by shifting the origin away from the inversion center, a topological invariant that would be trivial based solely on the OAI parity eigenvalues $\xi_{m}(\bs{k}_{\alpha})$ will be classified as non-trivial. Therefore, incorporating $\xi^{\text{A}}_{m}(\bs{k}_{\alpha})$ into Eq.~(1) of the main text, we obtain topological invariants independent of unit cell conventions.

\section{Numerical results: diamond and germanium}
\begin{figure*}
    \centering
    \includegraphics[width=1\linewidth]{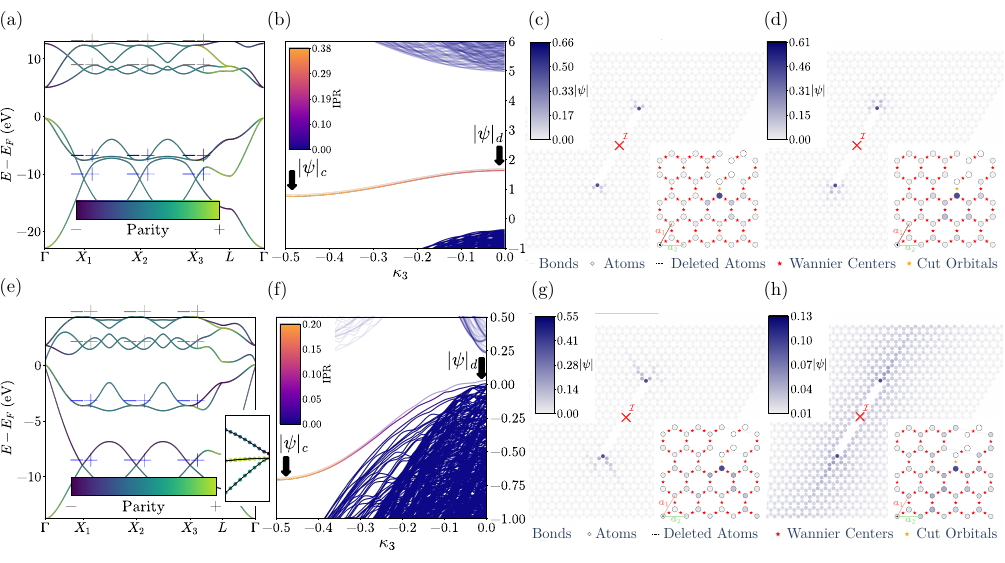}
    \caption{Tight-binding simulations of 3D diamond and germanium. (a) Energy dispersion relation for diamond, when following the path shown in the main text's Fig. 2(a), with associated inversion parity eigenvalues at TRIM points. (b) Corresponding diamond band structure, with energies plotted against the only translationally-invariant direction $\kappa_3$, which is along $\bs{b}_3$. (c) and (d) Corresponding real-space wavefunction $|\psi|$ plots for polarization states at $\kappa_3=-1/2$ and $\kappa_3=0$, respectively. 
    (e) Energy dispersion relation for germanium, when following the same path as in (a). (f) Corresponding germanium band structure, as in (b) for diamond. (g) and (h) Germanium real-space wavefunction plots similar to (c) and (d). In (a) and (e), blue-colored eigenvalues are for occupied bands; gray-colored eigenvalues are for unoccupied bands.
    Two inversion-related polarization bands appear in the middle of the gap for both (b) and (f). Although in each case we shifted one of them slightly for visualization, these two bands overlap when the gap is large. We color-code the IPR for these bands. It is maximal at $\kappa_3 = -1/2$ and minimal at $\kappa_3 = 0$, see (c) and (d), and (g) and (h).}
    \label{fig:diamond_germanium_polarization_results}
\end{figure*}
We summarize the numerical results for diamond and germanium in Fig.~\ref{fig:diamond_germanium_polarization_results}.
In addition we calculate the weak SSH invariants for black phosphorene (BP) and silicon (and hence also diamond and germanium) here. 

For BP, using the inversion eigenvalues in Figs. 1(a) and (b) and Eq. (2) in the main text, we obtain
\begin{equation}\label{eq: nus_for_BP}
\begin{aligned}
{\nu^{\mathrm{SSH}}_{1}} &= \frac12(1-\delta_{M=(1,1)}\delta_{X=(1,0)})= 1,\\ 
{\nu^{\mathrm{SSH}}_{2}} &= \frac12(1-\delta_{M=(1,1)}\delta_{Y=(0,1)})= 1.\\ 
\end{aligned}
\end{equation}
On the other hand, using the inversion eigenvalues in Figs. 2(a) and (b) and Eq. (5) in the main text, we have
\begin{equation}\label{eq: nus_for_silicon}
\begin{aligned}
{\nu^{\mathrm{SSH}}_{2,3}} &= {\nu^{\mathrm{SSH}}_{3,2}} = \frac12 \left[1-\delta_{X_1=(0,1,1)}\delta_{L=(1,1,1)}\right] = 1,\\ 
{\nu^{\mathrm{SSH}}_{3,1}} &= {\nu^{\mathrm{SSH}}_{1,3}} = \frac12 \left[1-\delta_{X_2=(1,0,1)}\delta_{L=(1,1,1)}\right]= 1,\\ 
{\nu^{\mathrm{SSH}}_{2,1}} &=
{\nu^{\mathrm{SSH}}_{1,2}} = \frac12 \left[1-\delta_{X_3=(1,1,0)}\delta_{L=(1,1,1)}\right]= 1 \\
\end{aligned}
\end{equation}
for all three 3D semiconductors silicon, diamond, and germanium.
\section{Effect of different terminations and inversion-symmetry breaking elastic relaxations}
\begin{figure}
    \centering
    \includegraphics[width=1\linewidth]{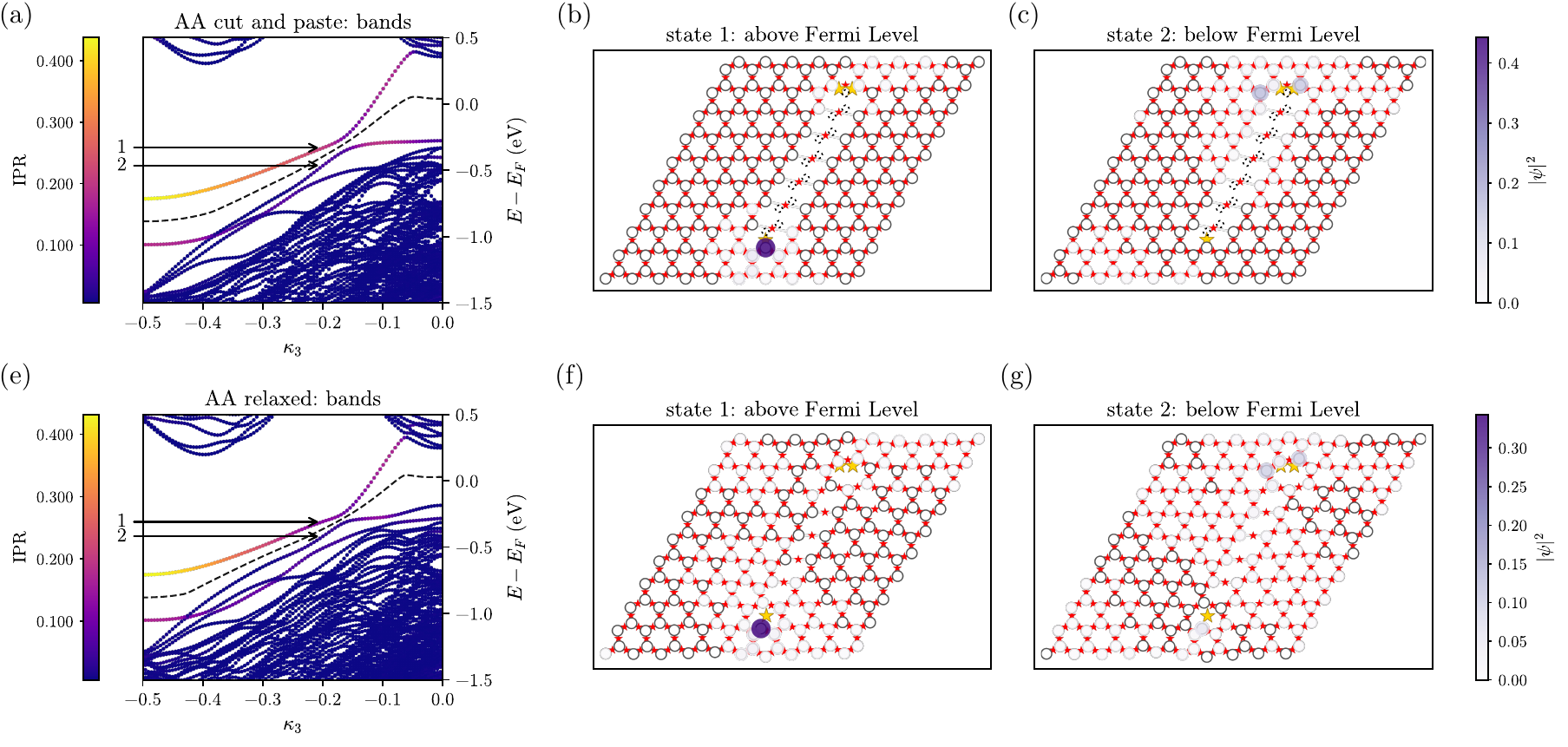}
    \caption{Tight-binding simulations for dislocations with AA termination in silicon. We plot both the energy spectra and density profiles of defect-localized states for an unrelaxed cut-and-paste (a-c) and relaxed (d-f) $10 \times 10$ unit-cell lattice. We ignore next-nearest neighbor hoppings. We plot the two states just above (labeled $1$) and below (labeled $2$) the momentum-resolved Fermi energy, as exemplified in panels (a) and (d) at $\kappa_3=-0.2$. The dashed black lines in these panels are visual aids to distinguish between the last filled and first unfilled band.}
    \label{fig:aa_results}
\end{figure}

To test the robustness of the localized states, we consider an alternative termination of the edge dislocation. In contrast to the ``AB'' atom termination used in the main text [see Fig.~2 there], we construct an ``AA'' atom termination by removing one additional atom at the upper end of the dislocation and reconnecting the corresponding bonds.
We then relax the reconstructed lattice using the Keating valence-force-field model for silicon, including both bond-stretching and bond-bending terms with force constants $\alpha = 2.965~\mathrm{eV}/\mathrm{\AA}^2$ and $\beta = 0.845~\mathrm{eV}/\mathrm{\AA}^2$, respectively~\cite{Keating1966}. After relaxation, the tight-binding hopping matrices are recalculated using the updated bond orientations and bond lengths. The distance dependence is taken from the Harrison scaling, $t(r)=t(r_0)(r_0/r)^2$~\cite{Harrison1980}. In this test, we keep only nearest-neighbor hoppings and neglect all next-nearest-neighbor terms.

The results are summarized in Fig.~\ref{fig:aa_results}. The ``AA'' termination explicitly breaks inversion symmetry, so the two localized bound states are no longer degenerate in the spectrum. The bound-state weight is also asymmetric between the two dislocation cores: the upper core carries less density, consistent with the local bond reconstruction restoring the cut Wannier-center structure there. After elastic relaxation, the two localized bands split further. The apparent reduction of the gap after relaxation can be understood from the increase of the average reconstructed bond length: this weakens the corresponding hopping amplitudes relative to the unrelaxed cut-and-paste Hamiltonian.
Overall, although changing the boundary termination and allowing inversion-asymmetric relaxations can reduce the polarization and visibility of the bound states, the dislocation modes remain in the gap and stay localized around $\kappa_3 = -0.5$.
\bibliography{refs}